\documentstyle[12pt,twoside,fleqn,espcrc1,epsfig]{article}

\newcommand \la{\raisebox{-.5ex}{$\stackrel{<}{\sim}$}}

\newcommand{\AmS}{{\protect\the\textfont2
  A\kern-.1667em\lower.5ex\hbox{M}\kern-.125emS}}

\title{Freeze-out from HBT and Coulomb Effects}

\author{H. Heiselberg\address{NORDITA, \\ 
        Blegdamsvej 17, DK-2100 Copenhagen O., Denmark}  }

\begin{document}
\maketitle

\begin{abstract}
The freeze-out of hot and dense hadronic matter formed in relativistic
nuclear collisions is probed by HBT interferometry of identical pions,
kaons, etc.  Coulomb repulsion/attraction of positive/negative
particles show up at small particle momenta and is also very sensitive
to the freeze-out conditions. The source sizes and times
freeze-out are extracted from $\pi^-/\pi^+$ spectra and HBT
radii and compared. 
\end{abstract}

\section{Coulomb Effects in Single Particle Spectra}

An asymmetry in the number of opposite charge pions has been observed
at intermediate energies and has recently also been
identified in heavy ion collisions at energies around 1 $A\cdot$GeV
\cite{Pelte}, 11.4 $A\cdot$GeV \cite{AGS}, and 158 $A\cdot$GeV
\cite{NA44}. The
ratio of negative to positive pions at low pion momenta is $\sim 3$ at
SIS energies, but only $\sim 1.6$ at AGS and SPS energies for the
central collisions of heavy nuclei as $Au$ or $Pb$. This effect can be 
explained by the Coulomb interaction between the produced pions
and the positive charge from the reaction partners.

At ultrarelativistic collision energies it is important to include the
rapid expansion of the system and its net charge as well as
retardation effects when considering the Coulomb effect on particles.
For rapid longitudinal Bjorken expansion we find (see \cite{Barz} for 
details and more elaborate calculations) that the
Coulomb field decrease with time as $\sim t^{-1}$ and is proportional
to the net charge (proton) rapidity distribution $dN^{ch}/dy$.  The
pions are assumed to freeze-out at time $\tau_f$ from a cylindrical
symmetric source of transverse size $R_f$.  On average a charged pion
of transverse momentum $p_{\perp,0}$ at freeze-out and final momentum
$p_\perp$ receives a momentum change or ``Coulomb kick'' $\pm p_c$ for
$\pi^\pm$ respectively
\begin{eqnarray}
  p_c \equiv |{\bf p}_\perp-{\bf p}_{\perp,0}|  &\simeq& 
   2e^2 \frac{dN^{ch}}{dy} \frac{1}{R_f} \, 
     \,. \label{pc} 
\end{eqnarray}
 The momentum change is only a slowly decreasing function
of the freeze-out time.
For example, $p_c$ varies $\pm$10\%
for $p_\perp\tau_f/m_\perp R_f$ in the range $0.5-0.8$.

The Coulomb effect on the transverse
particle distribution functions can be derived from particle conservation
$dN/d^2p_\perp = (p_{\perp,0}/p_\perp)\, dN_0/d^2p_{\perp,0}$.
The pion spectra are well reproduced by
$dN^0/d^2p_{\perp,0}\propto \exp(-m_{\perp,0}/T)$ with
$T\simeq150$ MeV at both AGS and SPS energies. 
Defining $m^\pm_\perp =\sqrt{m^2+(p_\perp\pm p_c)^2}$ the ratio is
\begin{eqnarray}
   \frac{\pi^-}{\pi^+} \equiv \frac{dN^-/d^2p_\perp}{dN^+/d^2p_\perp}
   = \langle\frac{\pi^-}{\pi^+}\rangle
    \frac{p_\perp+p_c}{p_\perp-p_c} 
     \exp\left(\frac{m^-_\perp-m^+_\perp}{T}\right)
\,.  \label{ratio3}
\end{eqnarray}

\begin{figure}[htb]
\begin{minipage}[t]{80mm}
\epsfig{figure=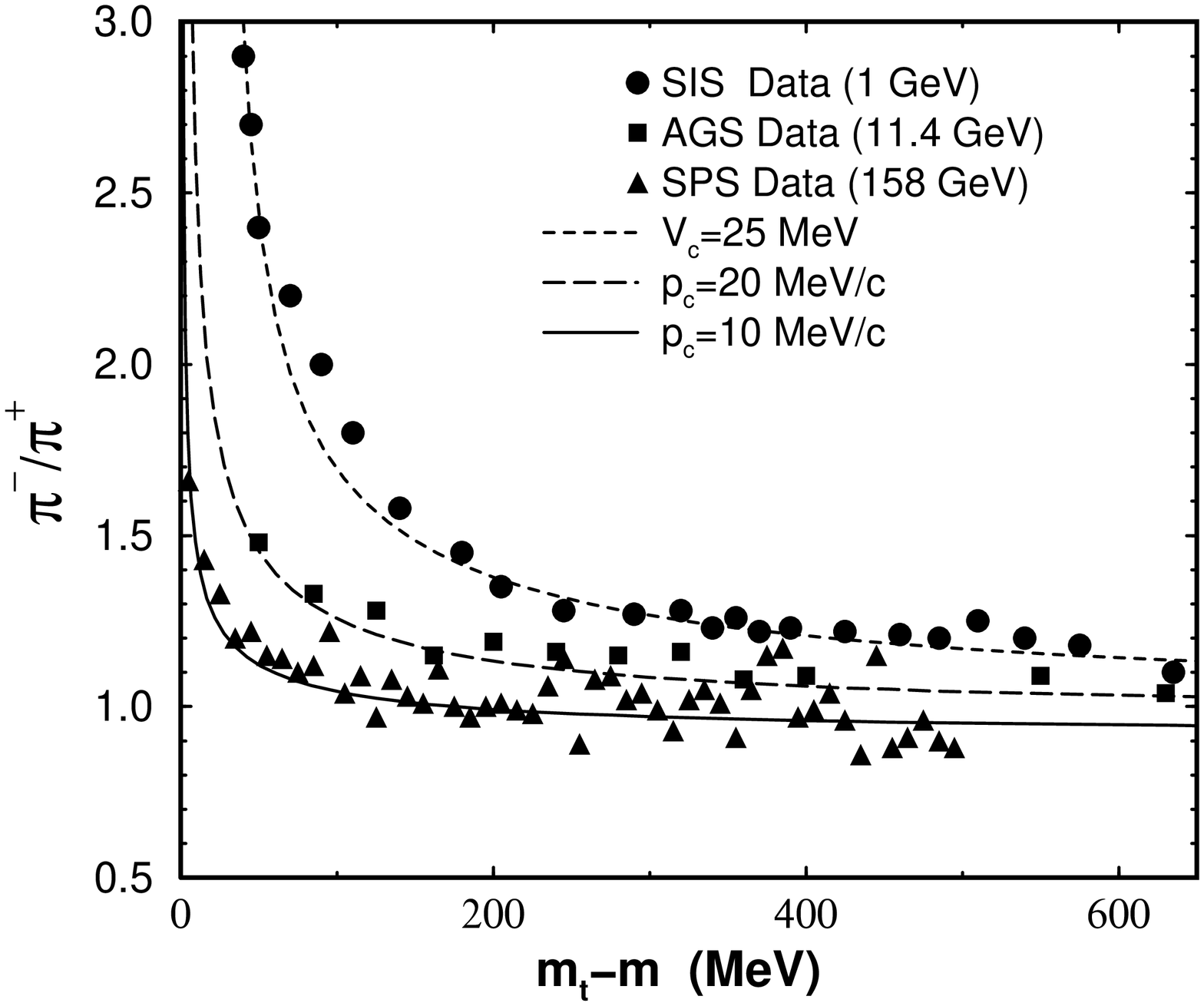,width=8cm,height=8cm,angle=-90}
\caption{$\pi^-/\pi^+$ ratios vs. $m_\perp$ at SIS,
AGS and SPS energies. Curves show model results (see text).}
\label{fig1}
\end{minipage}
\hspace{\fill}
\begin{minipage}[t]{75mm}
\epsfig{figure=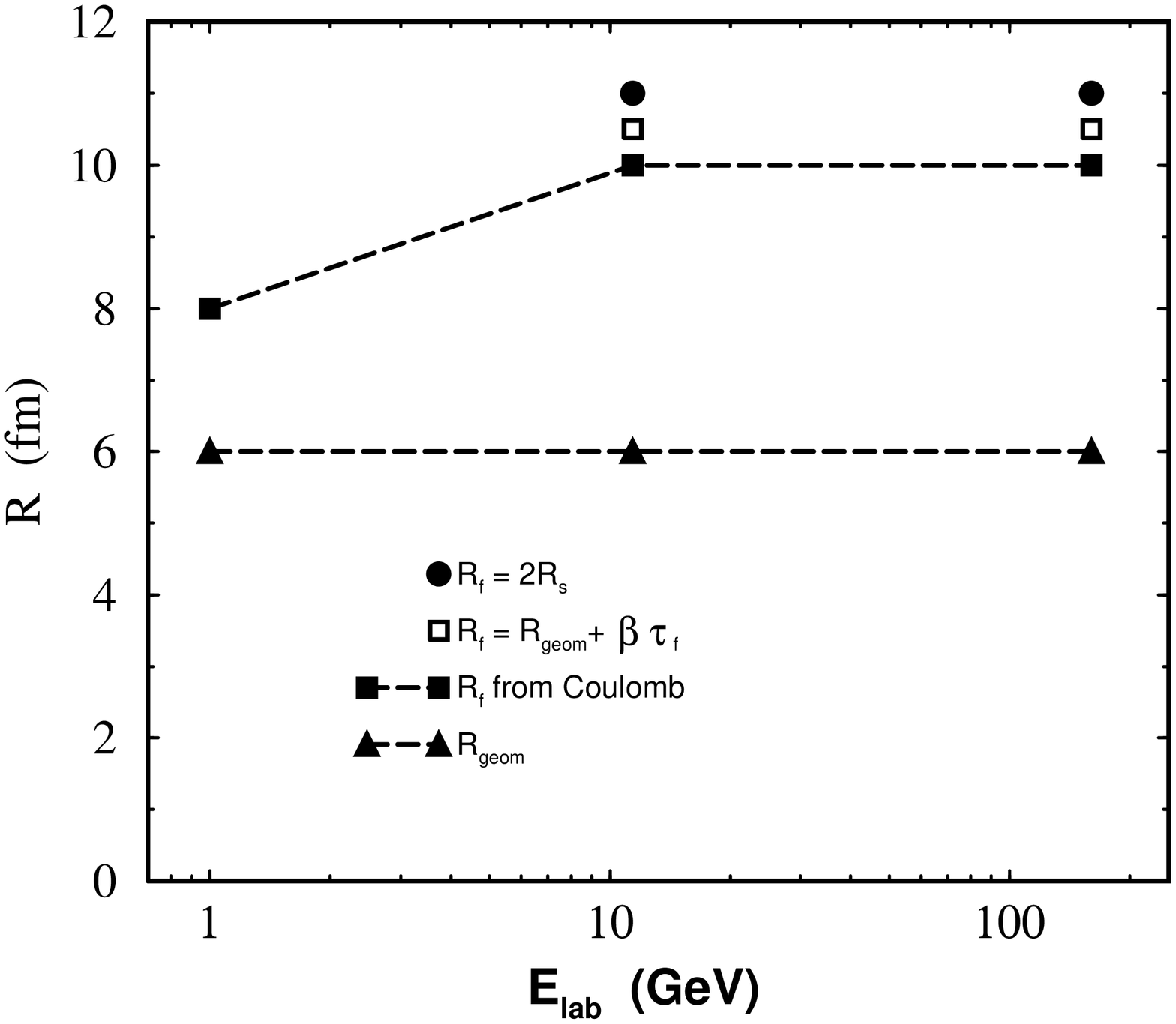,width=8cm,height=8cm,angle=-90}
\caption{Freeze-out radii extracted from Coulomb effects and HBT 
(see text) are larger than geometrical radii.}
\label{fig2}
\end{minipage}
\end{figure}

The pion ratio predicted by (\ref{ratio3}) is compared to AGS and SPS
data in Fig. \ref{fig1}. We use the experimental values for the total
pion ratio $\langle
\pi^-/\pi^+\rangle\simeq1.27$ and $1.05$ at AGS and SPS energies,
respectively. The experimental data is well reproduced with
$p_c^{AGS}\simeq 20$ MeV/c and $p_c^{SPS}\simeq 10$ MeV/c.  From the
measured proton rapidity distribution, $dN^p_{AGS}/dy\simeq 70$ and
$dN^p_{SPS}/dy\simeq 37$, and Eq.(\ref{pc}) we can now extract the
size of the systems at freeze-out: $R_f\simeq 10$ fm, at both
AGS and SPS.

The Coulomb effect is stronger at lower energies due to slow expansion
of the net charge.  A spherical and static charge of size $R$ changes
the particle energies by the Coulomb energy $V_c\simeq Ze^2/R$ (see,
e.g., \cite{Barz}) and leads to a larger $\pi^-/\pi^+$ ratio than for the
rapidly expanding case Eq. (\ref{pc}) as seen in Fig. (\ref{fig1}).
With $\langle\pi^-/\pi^+\rangle\simeq 1.9$ we extract 
from the data $V_c\simeq 27$MeV. With $Z=110$ for the $Au+Au$
collisions with 14\% centrality at SIS we obtain $R\simeq 8$fm.

The smaller charge in $S+S$ collision ($dN^p/dy\simeq 4$ at SPS)
leads to a smaller $\pi^-/\pi^+$ ratio that is compatible with
data \cite{NA44,Barz}.

Kaons are heavier than pions and are less affected by Coulomb
fields. Yet a small enhancement in the $K^-/K^+$ ratio at low $m_\perp$
is predicted \cite{Barz}. Preliminary data from NA44 \cite{NA44} shows a
flat $K^-/K^+$ ratio indicating that other physics is needed 
such as strong interactions counter-balancing the Coulomb fields.

\section{HBT Radii}

Bose-Einstein correlations (HBT) in relativistic heavy ion collisions
depend sensitively on the freeze-out condition. The source sizes are
measured in in the three directions, $R_s,R_o,R_l$, referred to as the
sideward, outward and longitudinal HBT radii. The theoretical
interpretation is, however, very model dependent. 

For a {\it transparent} source
the outward HBT radius is larger than the sideward
\cite{Csorgo,Heinz} 
\begin{eqnarray}
   R_o^2=R_s^2+\beta_o^2\sigma(\tau) \, , \label{R1}
\end{eqnarray} 
when opacity effects and transverse flow is ignored.
Here,  $\sigma(\tau)$ is the temporal fluctuations and contains 
the summed squares of the duration of emission, the average of short lived
resonance life-times and fluctuations in these \cite{HH}.
The difference between $R_o$ and $R_s$  
is due to the temporal extent
of the source during which the particles with outward velocity
$\beta_o=p_\perp/m_\perp$ travel on average a distance
$\beta_o\, \sigma(\tau)^{1/2}$ towards the detector.

In {\it opaque} sources and hydrodynamic 
models particles are only emitted from the surface contrary to the
transparent sources where particle freeze-out in volume.
By expanding for small relative momenta, $q_i\ll \hbar/R_i$, one can
express the HBT radii in terms of various fluctuations.
For very opaque sources of transverse size $R$, where the particle mean 
free path is short
$\lambda_{mfp}\ll R$, the HBT radii can be evaluated \cite{fluc}
($R$ is twice the commonly used gaussian radius parameter)
\begin{eqnarray}
   R_s^2 &=& \frac{1}{3} R^2
            \,-\,\frac{1}{6} \lambda_{mfp}^2  \,,\label{Rsf}\\
   R_o^2 &=& \left(\frac{2}{3}-(\frac{\pi}{4})^2\right) R^2
             \,+\, \beta_o^2 \sigma(\tau)
             \,+\, (\frac{7}{6}-\frac{\pi^2}{32}) \lambda_{mfp}^2  
              \,. \label{Rof}
\end{eqnarray}
In \cite{fluc} the HBT radii are also calculated with transverse flow and
moving surfaces.
Notice the different result for the 
transparent sources, for which $R_s=R/2$ and $R_o$ is
given by (\ref{R1}).
The opaque source has a much smaller transverse spatial extent
in the outward as in the sideward direction, 
$(2/3-(\pi/4)^2)\simeq 0.05\ll 1/3$, because the emission now take place
in a narrow surface layer of order the mean free path.

In relativistic heavy ion collisions
the outward and sideward HBT radii are measured to be similar
\cite{NA44,NA44QM,NA49} and in a few cases the outward is
even measured to be smaller than the sideward HBT radius 
\cite{NA44,NA44QM} contradicting Eq. (\ref{R1}). 
According to Eq. (\ref{R1}) this implies
that particles freeze-out suddenly, $\sigma(\tau)^{1/2}\ll R_i$, as in a
``flash'', in particular when resonance life-times are included.
However, both the opacity effect 
and transverse flow reduce $R_o$ more than $R_s$ and so it
is possible that $R_o<R_s$.
Present experimental results do not allow us to draw  conclusions
from the small difference between $R_o$ and $R_s$ since error bars are
of the same size. However, one cannot
use Eq. (\ref{R1}) to extract the duration of emission
since opacity and transverse flow can have significant effects.
Wiedemann et al. \cite{Heinz} have analysed the NA49 data and
exclude opacity effects. However, in the NA49 data
$R_o>R_s$ \cite{NA49}. 
It would therefore be interesting to investigate whether opacity effects
are needed to describe the NA44 data for which $R_o\la R_s$.

\section{Summary}

The $\pi^-/\pi^+$ ratios at SIS, AGS and SPS energies can be explained
by Coulomb effects with freeze-out radii about twice the
geometrical radii of the colliding nuclei. 
Similar freeze-out radii can be obtained
from HBT analyses of $R_s$ and $R_l$. 
The following freeze-out radii were extracted (see Fig. \ref{fig2}):\\
{\it i)}  
    From $\pi^-/\pi^+$ ratio we obtained best fits for 
    the Coulomb kick $p_c\simeq 2e^2(dN^p/dy)/R_f$
    at AGS and SPS and $V_c=Ze^2/R$ at SIS energies. These values 
    allowed us to estimate the freeze-out radii.\\
{\it ii)}
    For a source of constant pion density the sharp
    cut-of radius is given by $R_f=2R_s$, where the sideward HBT radius 
    $R_s\simeq 5-6$ fm in central $Au$ or $Pb$ collisions at AGS 
    \cite{Barrette} and SPS
    \cite{NA44QM,NA49}.\\ 
{\it iii)}
    In a longitudinally expanding system with Bjorken scaling the
    freeze-out time is related to the longitudinal HBT radius by $\tau_f=
    R_l\, \sqrt{m_\perp/T}$.  From the AGS and SPS data $R_l\simeq 5.5$ fm
    \cite{Barrette,NA44QM} we estimate $\tau_f\simeq 8$ fm/c.  Assuming
    an initial geometrical radius $R_{geom}\simeq$6fm
    (for $\sim15$\% centrality) with 
    transverse flow $\beta\simeq0.5-0.6c$ \cite{NA44slopes}, we find a
    radius of $R_f=R_{geom}+\beta\tau_f\simeq 11-12$fm.\\
{\it iv)}
    At present it is difficult to extract information on freeze-out from
    $R_o$ since it is very model dependent.  Opacity and transverse flow
    effects reduce it whereas duration of emission and resonances increase
    it with respect to $R_s$.

We conclude that both freeze-out times extracted from the longitudinal
and transverse HBT radii are at AGS and SPS energies and within
experimental uncertainty {\it compatible} with
those found from $\pi^-/\pi^+$ ratios as seen from Fig. \ref{fig2}.
The freeze-out radii are substantially larger than the geometrical
radii $R_{geom}\simeq 6$fm for near central collisions $Au$ or $Pb$
nuclei indicating that expansion takes place before freeze-out.

 Acknowledgements to my collaborators Barz, Bondorf, G\aa
rdh\o je and Vischer.

\end{document}